\documentclass[reprint,amsmath,amssymb,aps,prl,superscriptaddress]{revtex4-2}

\usepackage{graphicx}
\usepackage{color}
\usepackage{physics}
\usepackage{bm}
\usepackage[breaklinks=true,colorlinks,citecolor=blue,linkcolor=blue,urlcolor=blue]{hyperref}

\def\be {\begin{equation}}
\def\ee {\end{equation}}
\def\bea{\begin{eqnarray}}
\def\eea{\end{eqnarray}}

\begin{document}

\title{Nonlinear Valley Hall Effect}
\author{Kamal Das}
    \email{daskamal457@gmail.com}
    \affiliation{Department of Physics, Indian Institute of Technology, Kanpur-208016, India}
    \affiliation{Department of Condensed Matter Physics, Weizmann Institute of Science, Rehovot 7610001, Israel}
\author{Koushik Ghorai}
    \affiliation{Department of Physics, Indian Institute of Technology, Kanpur-208016, India}
\author{Dimitrie Culcer}
	\affiliation{School of Physics, The University of New South Wales, Sydney 2052, Australia}
    \affiliation{ARC Centre of Excellence in Future Low-Energy Electronics Technologies, The University of New South Wales, Sydney 2052, Australia}
\author{Amit Agarwal}
    \email{amitag@iitk.ac.in}
    \affiliation{Department of Physics, Indian Institute of Technology, Kanpur-208016, India}

\begin{abstract}

The valley Hall effect arises from valley contrasting Berry curvature and requires inversion symmetry breaking. Here, we propose a nonlinear mechanism to generate a valley Hall current in systems with both inversion and time-reversal symmetry, where the linear and second-order charge Hall currents vanish along with the linear valley Hall current. 
We show that a second-order valley Hall signal emerges from the electric field correction to the Berry curvature, provided a valley-contrasting anisotropic dispersion is engineered. We demonstrate the nonlinear valley Hall effect in tilted massless Dirac fermions in strained graphene and organic semiconductors. Our work opens up the possibility of controlling the valley degree of freedom in inversion symmetric systems via nonlinear valleytronics. 
 
\end{abstract}

\maketitle

{\it Introduction:---} 
Comprehending and controlling diverse degrees of freedom in quantum materials not only enriches our understanding of fundamental physics, it also paves the way for novel applications. For example, the fields of electronics and spintronics emerged from understanding and exploiting the charge and spin degree of freedom~\cite{rigor_RMP2004_spin,hirohata_SD2020_spin}, respectively. Beyond these, the valley degree of freedom has attracted significant attention giving birth to the field of  valleytronics~\cite{yamamoto_JPSJ2015_valley,schaibley_NRM2016_valley,vitale_small2018_valley}. Valleys are the degenerate energy extrema of the electronic bands in momentum space. Valleys become well-defined degrees of freedom when they are well separated in momentum space with negligible inter-valley scattering. The primary focus of valleytronics is to control and manipulate the valley degree of freedom by using   electrical~\cite{xiao_PRL2007_valley,gorbachev_S2014_detect,sui_NP2015_gate,shimazaki_NP2015_gen}, optical~\cite{yao_PRB2008_valley,cao_NC2012_valley,mak_NN2012_control,zeng_NN2012_valley}, and magnetic~\cite{li_PRL2014_valley,macneill_PRL2015_break,srivastava_NP2015_valley} means.

The linear valley Hall effect (VHE), introduced by Xiao et al.~\cite{xiao_PRL2007_valley}, enables electrical control and manipulation of the valley degree of freedom. VHE is the accumulation of electrons with opposite valley indices on opposite sides of a sample, transverse to the direction of the applied electric field. It is induced by the valley contrasting Berry curvature and anomalous Hall velocity in materials with broken inversion symmetry. It offers a way to probe the Berry curvature in time-reversal preserving systems where the anomalous charge Hall response vanishes. VHE has been measured via non-local resistance measurements in hexagonal graphene superlattices \cite{gorbachev_S2014_detect, sui_NP2015_gate, shimazaki_NP2015_gen} and in monolayer transition metal dichalcogenides~\cite{mak_S2014_the}.
However, in systems with inversion and time-reversal symmetry, the Berry curvature vanishes for each point in momentum space, leading to the absence of VHE in such systems. This raises a fundamental question. How can we probe and manipulate the valley degree of freedom in nonmagnetic and inversion symmetric materials by electrical means?

\begin{figure}[t]
\centering
\includegraphics[width = 0.9\linewidth]{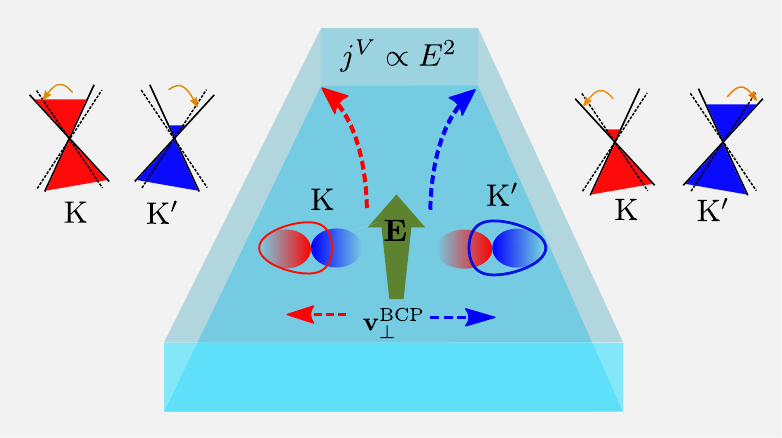}
\caption{Schematic of the nonlinear valley Hall effect (NVHE). A longitudinal electric field induces a nonlinear valley Hall current ($j^{\rm V} \propto E^2$). Due to the Berry connection polarizability (BCP) induced Hall velocity, electrons near the Fermi surface with different valley degrees of freedom accumulate on opposite sides of the sample. The elliptical density shows the BCP induced Hall velocity distribution, and the contours show the Fermi surfaces. The NVHE is finite in inversion and time reversal symmetric systems such as tilted gapless graphene, where the linear valley response and linear and nonlinear charge responses vanish.}
\label{fig:fig1}
\end{figure}

Here, we demonstrate that the nonlinear valley Hall effect (NVHE) can probe and manipulate the valley degree of freedom in inversion and time-reversal symmetric systems. We show that materials with both fundamental symmetries exhibit a finite NVHE that is second-order in the electric field (see Fig.~\ref{fig:fig1}), provided their band dispersion is anisotropic and valley contrasting~\cite{li_2Dmat2021_reveal,wild_arxiv2023_optical,zhang_arxiv2023_tunneling}. Such a dispersion can be engineered by reducing crystalline symmetries while retaining inversion and time-reversal symmetry, via strain or other means. The NVHE originates from the electric field-induced correction to the Berry curvature, which results in a nonlinear anomalous Hall velocity. Since the electric field correction is determined by the Berry connection polarizability (BCP)~\cite{gao_PRL2014_field, gao_FP2019_semi}, the NVHE can be used to investigate the quantum metric in systems with both fundamental symmetries.

We demonstrate that the charge carriers of opposite valleys carry opposite signs of the electric field-induced orbital magnetic moment (OMM). Thus, the NVHE separates carriers with opposite electric field-induced OMM, even though the total field-induced orbital magnetization vanishes in the system. Additionally, we highlight the \textit{intrinsic} (scattering-time independent) nature of the NVHE and show that the dominant part of the disorder-induced extrinsic contributions in NVHE vanish for tilted Dirac Hamiltonian. Experimentally, the NVHE can be probed via nonlocal resistance measurements using the scaling of the nonlocal resistance with the longitudinal ohmic resistivity, which is different from the case of the linear VHE. As specific examples, we demonstrate the existence of NVHE in strained graphene, which hosts tilted massless Dirac fermions, and in the organic semiconductor $\alpha ({\rm BEDT-TTF})_2 {\rm I}_3$.

{\it Theory of nonlinear valley Hall effect:---}  
The linear valley Hall current ($j_a^{\rm V}$) is defined in terms of the linear valley Hall conductivity ($\sigma_{ab}^{\rm V}$) as $j_a^{\rm V}= \sigma_{ab}^{\rm V}E_b$. Here, $j_a^{\rm V}=j_a^{\rm K}-j_a^{{\rm K}'}$ and $\sigma_{ab}^{\rm V}=\sigma_{ab}^{\rm K}-\sigma_{ab}^{{\rm K}'}$, with the distinct  valleys specified by ${\rm K}$ and ${\rm K'}$. Under time reversal, the valley current ($j_a^{\rm K} \to -j_a^{{\rm K}'}$ and $j_a^{{\rm K}'} \to -j_a^{\rm K}$) and the electric field do not change sign. However, under space inversion, the valley current remains unchanged but the electric field changes sign. These symmetry considerations force the  linear valley Hall current to vanish in an inversion symmetric system. In this background, we introduce the notion of NVHE originating from the nonlinear charge response specified by $j_a^{(2)}=\chi_{a;bc}E_b E_c$. We define the valley-resolved nonlinear charge conductivity as, $j_a^{{\rm V}(2)}=\chi_{a;bc}^{\rm NLV} E_b E_c$, where  
\be \label{NLVC}
\chi_{a;bc}^{\rm NLV}=\chi_{a;bc}^{\rm K}-\chi_{a;bc}^{{\rm K}'}~.
\ee
In an inversion symmetric system, the second-order nonlinear charge current vanishes, and the dominant valley response is the nonlinear valley Hall current~\footnote{This is reminiscent of the linear valley Hall current in time-reversal symmetric systems, where the anomalous charge Hall response vanishes.}.

The origin of the NVHE can be understood from the semiclassical electron dynamics by incorporating corrections up to second-order in the electric field~\cite{sundaram_PRB1999_wave,gao_PRL2014_field}. In the nonlinear regime, the anomalous Hall velocity $\bm{v}_{\rm AHE} = (e/\hbar) \bm {E} \times \bm {\Omega}_n$, arising from the interband coherence,  gets a nonlinear correction induced by the external electric field. The modification yields $\bm{v}_{\rm AHE} \to {\bm v}^{\rm E}_{\rm AHE} = \frac{e}{\hbar}{\bm E} \times ({\bm \Omega}_n  + {\bm \Omega}_n^{\rm E})$ where ${\bm \Omega}_n={\bm \nabla}_{\bm k} \times { \bm {\mathcal A}}_n$ is the usual Berry curvature and ${\bm \Omega}_n^{\rm E}$ is an electric field induced correction given by
\be 
{\bm \Omega}^{\rm E}_n= {\bm \nabla}_{\bm k} \times { \bm {\mathcal A}}^{\rm E}_n;~~~{\mathcal A}^{\rm E}_{n,a} = 2e \sum_{m \neq n}  \frac{{\rm Re}[{\mathcal R}^a_{nm} {\mathcal R}^b_{mn}]}{\epsilon_n - \epsilon_m} E^b~.
\ee
Here, ${ \bm {\mathcal A}}^{\rm E}_n$ is an electric field-induced correction to the Berry connection. The quantity ${\mathcal R}^a_{nm} $ represents the band-resolved Berry connection, and $\epsilon_n$ is the eigenvalue of the unperturbed Hamiltonian. To facilitate interpretation, the correction to the Berry connection is expressed as ${\mathcal A}^{\rm E}_{n,a} = \tilde {\mathcal G}_n^{ab}E_b$, where the BCP, $\tilde {\mathcal G}_n^{ab}$, is defined as~\cite{lai_NN2021_third}
\be 
\tilde {\mathcal G}_n^{ab} = 2e \sum_{m \neq n} \frac{{\rm Re}[{\mathcal R}^a_{nm} {\mathcal R}^b_{mn}]}{\epsilon_n - \epsilon_m}~.
\ee
We emphasize that the BCP contains the band resolved quantum metric~\cite{gao_PRL2020_tun,bhalla_PRL2022_reso}, ${\mathcal G}^{ab}_{mn}={\rm Re}[{\mathcal R}^a_{nm} {\mathcal R}^b_{mn}]$, which quantifies the distance between wave functions in the parameter space.

In the second order in the electric field, two Hall responses originate from the Berry curvature and the electric field induced correction to it. The anomalous Hall velocity combined  with the linear non-equilibrium distribution function gives rise to the Berry curvature dipole (BCD) induced nonlinear Hall conductivity ~\cite{sodemann_PRL2015_quant, sinha_NP2022_berry}. It is given by $\chi_{a;bc}^{\rm BCD} = g_s \frac{e^3\tau}{\hbar^2}~ \epsilon_{abd} \sum_{n, \bm k} \left( \partial_c \Omega_{n}^{d} \right) f_n$. Here $f_n$ is the equilibrium Fermi-Dirac distribution function, $\tau$ is quasiparticle scattering time, $\epsilon_{abd}$ is the anti-symmetric tensor of rank three and $g_s=2$ accounts for the spin degeneracy factor. See Sec.~S1 of the SM~\footnote{\href{https://www.dropbox.com/scl/fi/if8pyzlq804a088g6ykvj/SM_NVHE_arXiv_final.pdf?rlkey=mmnhxia5u64xtjst4j0besq1g&dl=0}{The Supplemental material} discusses, i) Theory of second-order nonlinear conductivity, ii) Theory of electric field-induced orbital magnetic moment iii) Difference between linear and nonlinear valley Hall effect, iv) Calculation for massive tilted Dirac system, v) NVHE without tilt vi) Extrinsic contributions to the valley Hall effect, vii) Experimental probe of nonlinear valley Hall effect, and viii) Tight-binding model calculations for strained graphene and organic conductor ix) Symmetries and materials for NVHE. } for a detailed derivation. The electric field correction to the Berry curvature gives rise to an intrinsic nonlinear Hall current specified by $j_a = - \frac{e^2}{\hbar} \epsilon_{abl} \sum_{n, \bm k}E_b ({\bm \nabla}_{\bm k} \times { \bm {\mathcal A}}^{\rm E}_n)_l f_n$. The corresponding intrinsic nonlinear conductivity is given by~\cite{wang_PRL2021_intrin,liu_PRL2021_intrin,gao_science2023_quantum,mazzola_NL2023_discov,wang_PRL2023_intrin,lahiri_arxiv2023_intrinsic}, 
\be \label{int_BCP}
\chi_{a;bc}^{\rm BCP} = - g_s \frac{e^2}{2\hbar}  \sum_{n, \bm k} \left( 2\partial_a \tilde {\mathcal G}_n^{bc} - \partial_b \tilde {\mathcal G}_n^{ac} - \partial_c \tilde {\mathcal G}_n^{ab} \right) f_n~.
 \ee
In the presence of inversion symmetry, both these contributions to the Hall charge current vanish. Below, we show that although the charge current vanishes in systems with both time reversal and inversion symmetries, the valley-resolved intrinsic contribution survives and generates a finite NVHE.

To understand this, we note that in a spinless system the Berry curvature satisfies ${\bm \Omega}_n(-{\bm k})=-{\bm \Omega}_n({\bm k})$ in presence of time reversal symmetry, and ${\bm \Omega}_n(-{\bm k})={\bm \Omega}_n({\bm k})$ in presence of inversion symmetry. As a consequence, when both symmetries are present, ${\bm \Omega}_n({\bm k})=0$ at each point in the momentum space. This forces the linear VHE to vanish. The contribution from the Berry curvature dipole also vanishes for each valley. However, the intrinsic contribution originating from Eq.~\eqref{int_BCP} for each valley, does not vanish and can be finite. The BCP transforms as {${\tilde{\mathcal G}}^{ab}_{n}(-{\bm k}) = {\tilde{\mathcal G}}^{ab}_{n}({\bm k})$} under the action of both the space-inversion and time-reversal. Therefore, unlike the Berry curvature, it can be finite at each point in momentum space in the simultaneous presence of both these symmetries. As a consequence, we can have a nontrivial NVHE. This opens up the potential for controlling the valley degree of freedom in systems with both fundamental symmetries. This is the main highlight of this paper. The subtle point is that even though both the symmetries are preserved globally, they must be broken in the individual valleys (or locally) to obtain a finite NVHE.

{\it Electric field-induced orbital magnetic moment:---} 
Similar to the spin Hall effect that separates opposite spins, the linear VHE segregates carriers with valley contrasting OMM in real space by pushing the oppositely OMM polarized carriers to different transverse edges. However, in the presence of both fundamental symmetries, the OMM vanishes in the whole Brillouin zone. This raises a  question. What physical quantity distinguishes the carriers separated by the NVHE in real space?

We find that the electric field can induce a correction to the OMM.
The $a$-th component of the field-induced OMM for $n$-th band carriers is given by~\cite{xiao_PRB2021_thermo,xiao_PRB2021_adia}
\be \label{mE}
{m_
a^{\rm E}(n)}=\sum_{l \neq n} \biggl[2e \dfrac{\textrm{Re} [{\mathcal{M}}_{ln}^{a}\mathcal R_{nl}^d] }{\varepsilon_n - \varepsilon_l} + \dfrac{e^2}{2\hbar}\epsilon_{abc}(\partial_b \mathcal G_{ln}^{cd})\biggr] E_d~.
\ee
Here, $\bm{\mathcal M}_{ln} = (e/2)\sum_{j\neq n}(\bm{v}_{lj}+\bm{v}_{n}\delta_{lj})\times \bm{\mathcal R}_{jn}$ is the interband OMM with $\bm{v}_{lj}$ the matrix element of the velocity operator $\hat v^a = (1/\hbar)\partial_a \mathcal H$, and we have defined ${\bm v}_{n} \equiv {\bm v}_{nn}$. The NVHE separates carriers with valley-contrasting field-induced OMM given in Eq.~\eqref{mE}. We find that materials possessing a valley-contrasting orbital magnetization show the NVHE. See Fig.~S1, and Sec.~S2 and S3 of the SM~\cite{Note2} for a detailed discussion. Below, we show this explicitly for strained graphene and in organic conductors.

{\it Nonlinear valley Hall effect in tilted Dirac systems:---} 
We now demonstrate the NVHE in a two-dimensional tilted Dirac system with two valleys~\footnote{The tilt can arise in graphene with uniaxial strain (see Sec.~S4 of SM~\cite{Note2} for details)}. The system is described by the Hamiltonian, 
\be \label{ham}
{\mathcal H}(s) = \hbar v_F  (s k_x \sigma_x +   k_y\sigma_y) +   s \hbar v_{t} k_x \sigma_0~.
\ee
Here, $v_F$ is the Fermi velocity, and $\sigma_i$'s are 
the Pauli matrices representing the sub-lattice degree of freedom. In Eq.~\eqref{ham}, $s = \pm 1$ denotes the valley index, and the ${\bm k}$ for each valley is measured from the two Dirac points located at K or K' point. The $v_t$ term tilts the Dirac cone along the $k_x$-axis in opposite directions for the two valleys. The tilt term breaks the space-inversion and time-reversal symmetries for each valley as $\epsilon(-{\bm k}) \neq \epsilon({\bm k})$. However, the pair of oppositely tilted Dirac nodes preserves both the fundamental symmetries globally.

The energy dispersion for this two-band model is given by {$\epsilon_\lambda = s\hbar v_t k_x +\lambda \hbar v_F k$}, with $k = \left(k_x^2 + k_y^2\right)^{1/2}$ and $\lambda = \pm 1$ denotes the band index. The $z$ component of the Berry curvature for this system vanishes in the entire Brillouin zone. We calculate the elements of the quantum metric to be $\mathcal G^{xx}_{cv}=\mathcal G^{xx}_{vc}$ and $\mathcal G^{yy}_{cv}=\mathcal G^{yy}_{vc}$, with  
{\be  \label{qntm_met}
{\mathcal G}^{xx}_{cv}= \frac{k_y^2}{4k^4},~
{\mathcal G}^{yy}_{cv}= \frac{k_x^2}{4 k^4},~{\rm and}~
{\mathcal G}^{xy}_{cv}={ \mathcal G}^{xy}_{vc} = - \frac{ k_x k_y}{4 k^4}~. 
\ee}%
We note that the quantum metric is independent of the tilt velocity and the valley index. Furthermore, in contrast to the Berry curvature, the sub-lattice (or inversion) symmetry-breaking gap parameter is not needed to have a finite quantum metric.

We calculate the valley and band-resolved nonlinear Hall conductivity to be $\chi^{\rm BCP}_{x;yy} (s,\lambda) = -s \lambda \frac{e^3 v_t}{4 \pi \mu^2}.$ This is in the limit of small tilt velocity, $v_t \ll v_F$, for the chemical potential $\mu$ measured from the Dirac point. The valley dependence (captured by $s = \pm 1$) in the valley resolved nonlinear intrinsic Hall current leads to the accumulation of the electrons with opposite valley index on the opposite side of the samples. 
This results in NVH conductivity, 
\be \label{sigma_valley}
\chi^{\rm NLV}_{x;yy}  = - \lambda\frac{e^3 v_t}{2 \pi \mu^2}~.
\ee
The NVHE depends on the tilt velocity and vanishes as $v_t \to 0$. This highlights that for each valley, the rotation symmetry (in the continuum model) or the $C_3$ symmetry (in the lattice model) has to be broken to have a finite NVHE. In experiments, this is achieved via strain or substrate effects. See Sec.~S4 of the SM~\cite{Note2} for details. To get an estimation of the strength of the NVHE, we define the NVH angle as $\theta^{\rm NLV}=j^{{\rm V}(2)}/j_{\rm L}$, where $j_{\rm L}$ is the linear longitudinal current. {For the typical value of model parameters: $v_t=0.3v_F$ with $v_F=10^6 $ m/s, $\mu=0.1$ eV, $\tau=1 $ ps, and $E\sim 1 {\rm V}/\mu {\rm m}$, we find that the NVH angle is $\theta^{\rm NLV} \approx 0.006 \%$. See Sec. S4 (C) in SM~\cite{Note2} for details.}  The NVHE with a different type of band anisotropy and trigonal warping has been demonstrated in Sec.~S5 in SM~\cite{Note2}.

\begin{figure}[t]
\centering
\includegraphics[width = \linewidth]{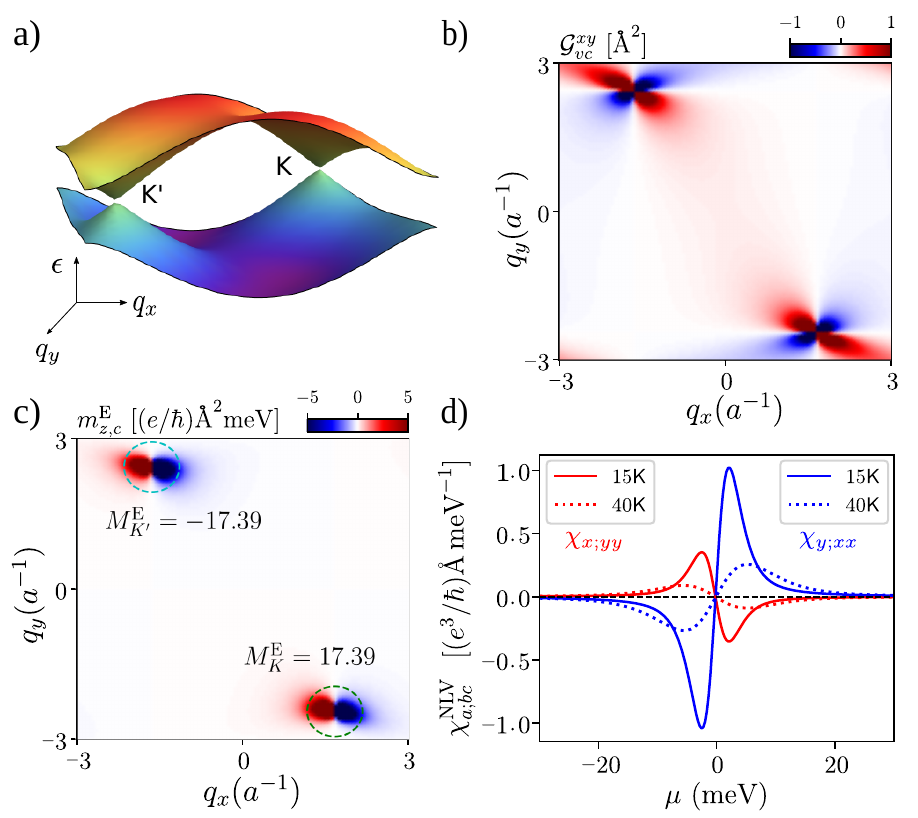}
\caption{(a) The electronic dispersion showing the two valleys of organic conductor $\alpha ({\rm BEDT-TTF})_2 {\rm I}_3$. The momentum space distribution of (b) the band-resolved quantum metric, and (c) the electric field-induced orbital magnetic moment for the conduction band. We have highlighted the valley contrasting orbital magnetization ($M^{\rm E}$) of the valleys. (d) Nonlinear valley Hall conductivities induced by the Berry connection polarizability for two representative temperatures: $15$K and $40$K. We have offset the energy axis by $-92~$meV so that $\mu =0$ represents the band touching point. We have used the following Hamiltonian parameters: $\{t_1, t'_1, t_2, t'_2, t_{\rm nn} \} = \{36,-86,-24,-77,-60\}$ meV for our numerical calculations.
\label{fig:fig2}}
\end{figure}

We can interpret our result of Eq.~\eqref{sigma_valley} as an accumulation of carriers with valley contrasting electric field-induced OMM along the edge of the sample. For ${\bm E}=E_y \hat {\bm y}$, using Eq.~\eqref{qntm_met} in Eq.~\eqref{mE} we calculate, 
\be 
{m_{z}^{\rm E}(s,\lambda)} = -\frac{e^2}{8\hbar}\frac{k_x}{k^4}\left(1-2\lambda s\frac{v_t}{v_F}\frac{k_x}{k} \right)E_y~.
\ee
Integrating this equation, we find that the total electric field induced OMM for the two valleys is equal in magnitude with opposite signs, $M_a^{\rm E} (s,n)=\int_{K_s} m_a^{\rm E} (s,n) f_n \propto s$. Here, $\int_{K_s}$ represents the integration near the valleys. Going beyond the low energy description, we also demonstrate the opposite orbital magnetization of the two valleys for an organic conductor in Fig.~\ref{fig:fig2}(c), and in the tight-binding model of strained graphene in {Fig.~S7 (c)} of the SM~\cite{Note2}.

{\it Extrinsic contributions to the valley Hall effect:---} 
For the linear VHE, in addition to the Berry curvature-induced intrinsic contribution, there are also finite extrinsic contributions induced by asymmetric scattering due to disorder 
~\cite{sinitsyn_PRB2007_anom}. Such extrinsic contributions arise from the side-jump and skew-scattering mechanisms~\cite{glazov_PRL2020_skew,glazov_PRB2020_valley}. Intriguingly, these extrinsic contributions are known to sometimes partially compensate, if not cancel, the intrinsic contribution. This motivates us to explore extrinsic disorder-induced contributions to linear VHE, as well as  NVHE predicted in this paper. Although a detailed disorder calculation within the quantum kinetic theory ~\cite{atencia_PRB2023_disorder} is deferred to a future publication, we find that both the skew scattering and side-jump semiclassical contributions vanish in the linear and nonlinear regime for the Hamiltonian in Eq.~\eqref{ham}. 

The extrinsic side-jump contribution is determined by the positional shift $\delta {\bm r}_{l'l}$ caused by the side-jump process, which subsequently affects the side-jump velocity, ${\bm v}^{\rm sj}=\sum_{l'} \bar \omega_{ll'}^{\rm sy} \delta {\bm r}_{l'l}$. Here, $l$ and $l'$ represent the quantum state involved in scattering, and $\omega_{ll'}^{\rm sy}$ is the symmetric scattering probability. Interestingly, the side-jump positional shift is determined by the Berry curvature~\cite{sinitsyn_PRB2005_disorder,sinitsyn_PRB2007_anom}. Hence, in systems with both fundamental symmetries, the principle part of the side-jump contribution is expected to vanish along with the Berry curvature. See Sec.~S6 (B) of SM~\cite{Note2} for details. On the other hand, the contribution from the skew-scattering mechanism contribution is determined by the asymmetric scattering rate, and it is proportional to the cubic and quartic powers of the scattering potential. This asymmetric scattering also vanishes for the Hamiltonian in Eq.~\eqref{ham}~\cite{sinitsyn_PRB2007_anom, borunda_PRL2007_absence}. Therefore, in contrast to the linear VHE, the NVHE does not have any contribution from the disorder-induced asymmetric scattering mechanisms for the tilted Dirac model Hamiltonian.

{\it Experimental signature:---} 
The linear VHE has been experimentally probed via two types of measurements. The first approach is based on measuring the anomalous charge Hall response in valley-polarized systems~\cite{mak_S2014_the,cai_PRB2013_magnetic,wang_TRSC2021_valley}~\footnote{The valley polarization can be induced by illumination with a circularly polarized light~\cite{mak_S2014_the}, by applying a static magnetic field~\cite{cai_PRB2013_magnetic}, or by breaking crystalline symmetries~\cite{wang_TRSC2021_valley}.}.
The NVHE can be probed via a similar strategy of inducing valley polarization by similar or other means.

Another approach for detecting the VHE is via the nonlocal resistance measurement~\cite{gorbachev_S2014_detect,sui_NP2015_gate,shimazaki_NP2015_gen,sinha_NC2020_bulk}. In the presence of a long-range charge neutral valley Hall signal, the nonlocal resistance generated by the inverse VHE scales as the cubic power of the longitudinal charge resistivity, $(\rho^{\rm C}_{ xx})^3$. We show that a similar nonlocal Hall measurement setup can also measure the NVHE. However, the scaling of the nonlocal resistance generated by the NVHE will be different from that generated by other means. We calculate the nonlocal resistance originating from the NVHE to be
\be 
R_{\rm NL}^{\rm NVHE}(x) = \frac{W}{2l_{\rm v}}~(\rho^{\rm C}_{xx})^5~ ({\chi^{\rm V}_{x;yy})^2~j^2}~e^{-|x|/l_{\rm v}}~.
\ee
Here, $W$ is the width of the sample, $l_{\rm v}$ is the inter-valley scattering length, $j$ is the bias current density used for measurement, and $x$ is the distance from the nominal current path. The calculation details are presented in Sec.~S7 of the SM~\cite{Note2}. This has to be contrasted with the case when there is no VHE at all, for which we have 
$R_{\rm NL}^{\rm no-VH} \propto \rho^{\rm C}_{xx}$, and with the case for linear VHE for which we have $R_{\rm NL}^{\rm linear-VH} \propto (\rho^{\rm C}_{xx})^3 \sigma^{\rm V}_{xy}$. In contrast to these cases, the nonlocal resistance induced by NVHE is measurement current dependent, and it is proportional to $(\rho^{\rm C}_{xx})^5 j^2$. We predict that the NVHE manifests as a second harmonic signal in the nonlocal measurement setup, even though the system has inversion symmetry~\footnote{NVHE offers an additional experimental advantage over the linear VHE. The experimental detection of the linear VHE in graphene superlattices has been subject to some controversy. It is not clear whether the valley current is carried by the edge states~\cite{yao_PRL2009_edge,zhu_NC2017_edge,marmolejo_JPM2018_de,roche_JPM2022_have} or by the bulk states~\cite{lensky_PRL2015_topolo,gorbachev_S2014_detect,sui_NP2015_gate,shimazaki_NP2015_gen}. In contrast, the NVHE is free from this ambiguity since the valley current contributions are Fermi surface-dependent, implying that they originate from the bulk bands.}.

{\it Material realization:---} 
As a realistic material example, we calculate the NVHE for the organic conductor $\alpha ({\rm BEDT-TTF})_2 {\rm I}_3$~\cite{hirata_NC2016_obser,kobayashi_JPSJ2007_massless} {and graphene (graphene results are shown in Sec.~S8 of SM~\cite{Note2})} in presence of uniaxial strain~\cite{goerbig_PRB2008_tilted,choi_PRB2010_effects}. Under strain, $\alpha ({\rm BEDT-TTF})_2 {\rm I}_3$ supports a pair of oppositely tilted and gapless Dirac cones. It is described by the tight-binding Hamiltonian of the form ${\mathcal H}({\bm q})=[(h', h^*),(h, h')]$. The off-diagonal elements are given by
\be
h ({\bm q}) = 2 \Big[t_1 e^{iq^+/2} + t_1' e^{-iq^+/2} + t_2 e^{iq^-/2}+t_2' e^{-iq^-/2} \Big]~,
\ee
where $q^{\pm}=q_x \pm q_y$. Here, $t_i$ and $t_i' ~(i=1,2)$ are the nearest neighbor hopping amplitudes ({see Sec.~S8} of SM~\cite{Note2} for more details). The diagonal elements are given by $h'({\bm q})=2 t_{\rm nn} \cos q_y~$, where $t_{\rm nn}$ is the next neighbor hopping amplitude. We present the resulting electronic dispersion and the corresponding Dirac valleys in Fig.~\ref{fig:fig2}(a). In Fig.~\ref{fig:fig2}(b), we show the momentum space distribution of the quantum metric. In Fig.~\ref{fig:fig2}(c), we display the electric field-induced OMM which gives rise to valley orbital magnetization. Figure~\ref{fig:fig2}(d) illustrates the resulting NVH conductivity. The NVHE decreases as we increase the temperature showing the energy window in which the BCP tensor is finite around the band touching point.

{\it Symmetry and candidate materials:---} Nonmagnetic materials with an inversion center are the most suitable candidates for observing NVHE. Note that while the NVHE and intrinsic BCP Hall current~\cite{wang_PRL2021_intrin,liu_PRL2021_intrin} share the same physical origin, the fundamental symmetries for these two phenomena are mutually exclusive. In addition to the fundamental symmetries, crystalline symmetries like rotation and reflection near the valleys dictate various components of the NVHE tensor, as summarized in Table S1 of Sec.~S9 in the SM~\cite{Note2}. In gapless systems, the quantum metric peaks near the band crossings. Therefore, symmetry-protected gapless systems are likely to have large NVHE.
Furthermore, recent advancements in device technology have enabled the tunability of the bandgap in graphene superlattices through gate voltages~\cite{sui_NP2015_gate,shimazaki_NP2015_gen,sinha_NP2022_berry}, facilitating the creation of systems with gapless Dirac nodes on demand. An additional ingredient for inducing a significant NVHE signal is large anisotropy of the Fermi surface. Therefore, strain engineering to distort the Dirac nodes and introduce anisotropy~\cite{yang_PRB2018_effect} will help observe NVHE. Two-dimensional systems with anisotropy like borophene, the surface states of three-dimensional crystalline topological insulators~\cite{liu_PRB2013_two}, surface states of the crystalline topological insulator SnTe~\cite{tanaka_NP2012_expt} are good candidates for the observation of NVHE.

{\it Conclusion:---} 
In summary, we have predicted a new, NVHE in materials with spatial inversion symmetry. Our findings facilitate the control of the valley degree of freedom in centrosymmetric materials and offer several exciting possibilities. One interesting direction is  the field of valley-caloritronics, utilizing the possibility of the nonlinear valley Nernst efffect~\cite{dau_NC2019_the} 
and nonlinear valley thermal Hall effect. The NVHE can also be generalized to bosonic systems, with possibilities of predicting and observing the magnon contribution to the nonlinear valley thermal Hall effect~\cite{chen_PRB2022_magnon}. Additionally, nontrivial physics is likely to emerge in spin-orbit coupled systems where the spin-valley coupling has been shown to impact both the spin and valley Hall effect~\cite{xiao_PRL2012_coupled}.

{\it Acknowledgement:---} K.~D. thanks, Prof. Binghai Yan and Debottam Mandal, for the insightful discussion, and was supported by the Weizmann Institute of Science and the Koshland Foundation. K.~G. thanks to the MHRD, India for funding through the Prime Minister’s Research Fellowship (PMRF). A.~A. acknowledges the Department of Science and Technology of the Government of India for financial support via Project No.~DST/NM/TUE/QM-6/2019(G)-IIT Kanpur.

\bibliography{main.bib}

\end{document}